\documentclass[twocolumn,superscriptaddress,showpacs,preprintnumbers,prb]{revtex4}
\usepackage{graphicx}
\usepackage{amsmath} 
\usepackage{times}
\usepackage{epsfig}
\usepackage{multirow}
\usepackage{color}
\usepackage{ulem}   
\begin{document}

\def\salto{\vskip 1cm} \def\lag{\langle} \def\rag{\rangle}

\newcommand{\redit}[1]{\textcolor{red}{#1}}
\newcommand{\blueit}[1]{\textcolor{blue}{#1}}
\newcommand{\magit}[1]{\textcolor{magenta}{#1}}

\newcommand{\MSTD} {Materials Science and Technology Division, Oak Ridge
National Laboratory, Oak Ridge, TN 37831, USA}
\newcommand{\CNMS} {Center for Nanophase Materials Sciences, Oak Ridge National
 Laboratory, Oak Ridge, TN 37831, USA}
\newcommand{\LLNL} {Lawrence Livermore National Laboratory, Livermore, CA
94550, USA}

\newcommand{\NCSA} {National Center for Supercomputing Applications, University of Illinois at Urbana-Champaign, Urbana, IL 61801, USA}

\title{Simple Impurity Embedded in a Spherical Jellium:\\ 
Approximations of Density Functional Theory compared to\\ Quantum Monte Carlo Benchmarks}

\author{Michal Bajdich}           \affiliation {\MSTD}
\author{P. R. C. Kent}        \affiliation {\CNMS}
\author{Jeongnim Kim} \affiliation{\NCSA}
\author{Fernando A. Reboredo}       \affiliation {\MSTD}

\begin{abstract}
  We study the electronic structure of a spherical jellium in the
  presence of a central Gaussian impurity.
  We test how well the resulting inhomogeneity effects beyond spherical
  jellium are reproduced by several approximations of density
  functional theory (DFT).  Four rungs of Perdew's ladder of
  DFT functionals, namely local density approximation (LDA), 
  generalized gradient approximation (GGA), meta-GGA and orbital-dependent hybrid
  functionals are compared against our quantum Monte Carlo (QMC)
  benchmarks.  We identify several distinct transitions in the ground
  state of the system as the electronic occupation changes between
  delocalized and localized states. We examine the parameter space of
  realistic densities ($1 \le r_s\le 5$) and moderate depths of the
  Gaussian impurity ($Z<7$).  The selected 18 electron system (with
  closed-shell ground state) presents $1d \to 2s$ transitions while
  the 30 electron system (with open-shell ground state) exhibits $1f
  \to 2p$ transitions.  For the former system, the accuracy for the
  transitions is clearly improving with increasing sophistication of
  functionals with meta-GGA and hybrid functionals having only small
  deviations from QMC.  However, for the latter system, we find much
  larger differences for the underlying transitions between our pool
  of DFT functionals and QMC.  We attribute this failure to treatment
  of the exact exchange within these functionals.  Additionally, we
  amplify the inhomogeneity effects by creating the system with
  spherical shell which leads to even larger errors in DFT
  approximations.
\end{abstract}

\date{\today}

\maketitle
\section{Introduction}
Kohn--Sham density functional theory~\cite{Hohenberg64,Kohn65} (DFT)
is now arguably the most popular computational method in theoretical
condensed matter physics and materials science.  While the method is
exact in principle, in practice it is applied only with approximations
to the unknown, exact exchange-correlation functional.  The
conventional semilocal approximations often fail to describe the
structural, defect, and other properties of materials with strong
electron-electron correlations---a failure that is not only
quantitative but often qualitative. Indeed, even for less strongly
correlated materials where the method yields reasonable predictions,
further increases in accuracy are highly desired.
 
On the other hand, the quantum Monte Carlo (QMC) method allows direct
solution of 
the many-body problem of interacting electrons using stochastic
techniques~\cite{qmchistory1,qmchistory2,hammond,Mitas01}.  In fact,
the local density approximation is based on QMC calculations of the
homogeneous electron gas~\cite{DMC80} (HEG).  The only significant
source of systematic errors in the QMC method is the
fixed-node approximation~\cite{anderson,reynolds} to the Fermion
sign-problem. Using fixed-nodes, QMC has proven to been very effective
in providing high accuracy results for many real systems such as
molecules, clusters and solids with hundreds of valence electrons that
are within 1-3\% of
experiment~\cite{Mitas01,jeff_benchmark,nemec}. More recently, new
techniques have been developed to reduce the fixed-node
errors\cite{Umrigar07,SHDMC09,shdmcletter}, further improving the
accuracy of the technique.

The spherical jellium system has been extensively investigated 
as a model of large clusters
of simple metals (see, e.g., review~[\onlinecite{Brack93}]). It has been found to have pronounced 
shell structure with magic numbers $N=2,8,18,20,34,40,58,92,\ldots$ 
approximately corresponding to  fully-filled closed shell of each orbital momentum as $1s^2|1p^6|1d^{10}|2s^2|1f^{14}|2p^6|1g^{18}|2d^{10}1h^{22}3s^2$. 
A few benchmark QMC studies exist~\cite{Ballone92,Harris98,Sottile01} for the total 
and correlation energies of closed-shells at magic numbers with up to 106 electrons.   
These results have been later compared against several DFT approximations~\cite{Almeida02,Almeida08}. 
Surface correlation energies~\cite{Sottile01} and  
surface exchange-correlation energies~\cite{Almeida02,Almeida08} have been obtained using extrapolated values to infinite size. In general, the 
values obtained for correlation surface energies obtained with spheres are consistent with the latest DMC results for jellium slabs~\cite{Foulkes07} and other methods, 
but differ to some extent from the original DMC results from slab geometries~\cite{DMC96}. 

In this paper, we benchmark four levels of DFT approximations, 
following the ``Jacob's ladder'' metaphor of Perdew and co-workers~\cite{PerdewLadder,tpss,JPPerdewPrescriptionJCP2005}, 
by performing accurate quantum Monte Carlo calculations on the model of
the interacting electron gas subject to a central impurity potential.
The interacting electron gas is interpreted as a finite jellium sphere
with $N$ electrons of average density $r_s$.  An attractive spherical
Gaussian of tunable strength at the origin represents the impurity
potential.  The proposed system closely relates to an atom in a real
material while retaining a simplicity that is amenable to highly
accurate solution under a wide range of conditions (slowly- and
rapidly-varying density regions as well as a wide range of density
values).  The purpose of solving this simple model is to understand
the role of electronic correlations and exchange, to understand the features that more sophisticated 
functional must possess in order to describe, e.g., strongly correlated
materials, and to provide essentially exact benchmarks for testing new
DFT approximations.  Tabulated data of our results is given in Electronic Physics
Auxiliary Publication Service (EPAPS) Document No. [].

\section{Models with Spherical Symmetry}\label{model}
\subsection{Spherical Jellium Model}
One of the simplest ways to neutralize the negative charge of $N$ electrons
is to consider a sphere of positive charge $\rho_{B}$ of uniform density 
\begin{align}
\rho_{B}(r)=\left \{
  \begin{array}{ll}
    \frac{3}{4 \pi} r_s^{-3}, & r \le R_c \equiv r_s N^{1/3} \\
    0,  & r >  R_c,
  \end{array}\right.
\end{align}
with $r_s$ as adjustable parameter corresponding to a Wigner--Seitz radius in solids and $R_c$ as the sphere radius.
This is a basic description of the spherical jellium model. 
The external potential due to the positive background is then 
\begin{align}\label{eq:vext}
    V_{ext}^N(r)= \left\{
      \begin{array}{ll}
        -\frac{1}{2}\frac{N}{R_c}\left ( 3-\frac{r^2}{R_c^2}\right ), &  r \le R_c \\
        -\frac{N}{r}, & r >  R_c. 
      \end{array} \right.
\end{align}
The Hamiltonian of the electronic system (in atomic units) has the form
\begin{align}\label{Ham}
{\mathcal H}=-\frac{1}{2}\sum_i^N \nabla^2_i + \sum_i^N V_{ext}^N(r_i)+\frac{1}{2}\sum_{i,j\neq i}^N \frac{1}{|{\bf r}_i-{\bf r}_j|}+E_{self},
\end{align}
where we have introduced the constant Coulomb self energy of the positive background as
\begin{align}
E_{self}=\frac{3}{5}\frac{N^{5/3}}{r_s}.
\end{align}
The addition of $E_{self}$ ensures that in the thermodynamic limit the energy of jellium spheres approaches the energy of HEG.


\subsection{Spherical Jellium Model with a Gaussian Impurity}\label{sec3}
In order to further enhance the inhomogeneity effects in the jellium spheres, 
we propose the addition of the attractive Gaussian shaped impurity at the origin. 
The external potential is then modified as
\begin{align}\label{eq:vextwgauss}
    V_{ext}^N(r)= -Z\exp(-r^2/\sigma^2) + \left\{
      \begin{array}{ll}
        -\frac{1}{2}\frac{N}{R_c}\left ( 3-\frac{r^2}{R_c^2}\right ), &  r \le R_c \\
        -\frac{N}{r}, & r >  R_c 
      \end{array} \right.
\end{align}
where $Z$ and $\sigma$ represent the depth and the width of the added Gaussian. 
The Gaussian form for the impurity is not meant to exactly describe core levels of a real atom, 
but to create additional localization for valence electrons.  Moreover, using this model, we can also avoid 
the locality approximation required to evaluate conventional pseudopotentials~\cite{mitas91}. 
As a welcome consequence, the Gaussian potential also preserves 
the cusp-less property of single-particle orbitals at the origin. 
To summarize, by controlling the amount of localization at the
impurity, we can  alter the shape and occupation 
order of single-particle states. We note that a similar potential has been used in studies of 
hetero-atomic clusters of Refs.~\onlinecite{cohen87,Baladron88}.

\subsection{Spherical Jellium Shell  with a Gaussian Impurity}\label{sec4}
Another possibility to further amplify the inhomogeneity effects in the jellium 
is to create a spherical jellium shell. This can be achieved by combining 
external potentials [Eq.~(\ref{eq:vext})] of a larger system with $N+M$ electrons 
and smaller system with $M$ electrons to form
\begin{align}\label{eq:vextshell}
V_{shell}^{N}(r)= V_{ext}^{N+M}(r)-V_{ext}^{M}(r) -Z\exp(-r^2/\sigma^2), 
\end{align}
where integer variable $M \leq N$ controls the geometry of a shell [inner radius $R_c^M=r_sM^{1/3}$ 
and outer radius $R_c^{M+N}=r_s(M+N)^{1/3}$]. Note that in Eq.~(\ref{eq:vextshell}) 
the Gaussian impurity at the origin is also explicitly included.  
Above external potential closely resembles the  potential of hollow clusters~\cite{C20paper,hollowclusters09}.
The difference in our model is the addition of Gaussian impurity which  attracts electron density towards the origin. 
In turn, the enhanced inhomogeneity (i.e., charge separation between center and shell) provides even more severe 
test for the single-particle methods studied here.

\section{Methods}~\label{methods} The goal of this paper is to compare
the total energies and radial densities of equivalent states of the
spherical jellium systems in Hartree--Fock (HF), DFT and quantum Monte Carlo Methods.
Due to the spherical symmetry of the system and because the spin-orbit
interaction is not considered, the eigenstates of ${\mathcal H}$ must
also be eigenstates of the angular momentum operators $L^2$ and $M_L$,
spin operators $S^2$ and $M_S$.  The advantage of the spherical
symmetry is that rigorous upper-bound theorems apply for QMC and
ground state DFT can be generalized\cite{lundquist1976} for the lowest
energy state states of each value of $L$ and $S$. Therefore,
meaningful phase diagrams can be constructed with each method.

\subsection{Criteria used to select eigenstates of $L$ and $S$}
For benchmark purposes, we only need to compare equivalent states with
DFT and QMC. For comparison, we have selected states that are likely but
not guaranteed to be the minimum energy configurations.  Since we are
concerned with jellium densities of $1\le r_s \le 5$, our system can be
characterized as weakly interacting. In this regime it is a reasonable
to assume that, in analogy with atoms and 3D quantum dots~\cite{vorrath2003}, 
the filling of the orbitals within a single shell of spherical jellium follows Hund's rules,
i.e., for a given electron configuration, the state with maximum
multiplicity ($2S+1$) has the lowest energy, and for a given
multiplicity, the state with the largest value of $L$ is likely to 
have the lowest
energy. It is straightforward to construct the eigenfunction of
$^{2S+1}L$ symmetry (in the Russell--Saunders term symbol notation) as
the Slater determinant of single-particle orbitals with occupancies
chosen such that $M_L\equiv\sum_i m_i=L$ and $M_S\equiv\sum_i s_i=S$.
For $M_L\neq 0$, the determinants will be complex-valued, an issue
important in the QMC context and discussed further bellow.

The original second Hund's rule only applies to $3D$
spherically-symmetric systems, since it involves the total angular
momentum $L$.  There is debate, however, on the
existence and form of a second Hund's rule, with redefined shells,
in the case of parabolic models of
quantum dots in $2D$ at the highly correlated limit
(see, e.g., Refs.~\onlinecite{HaroldQD2007,UmrigarQD2000}). While that debate is
interesting, it is beyond the scope of
this paper to validate the Hund's rules in the range of parameters
explored with our $3D$ models.


\subsection{Single-particle Methods}
To obtain the quantities of interest within single-particle methods we
employ a non-relativistic atomic solver, originally 
implemented in Ref.~[\onlinecite{Vincent06}] and currently part of the
QMCPACK suite~\cite{qmcpack}, modified to handle pure HF method, 
LDA, GGA, meta-GGA and hybrid functionals. 
The single-particle orbitals are conveniently expressed as
radial-functions $R_{n,l}(r)$ multiplied by an angular-part given as
a spherical harmonic $Y_{l,m}(\theta,\phi)$.  This choice ensures well defined quantum numbers.

The radial equations [$R_{n,l}(r)$] 
are iteratively solved via Numerov algorithm on a large logarithmic grid ($\sim$ 5000 points) until the tight 
convergence criteria are met ($\Delta E_{tot}<10^{-7}$ and $\Delta E_{eig}<10^{-14}$).
For the later cases of the open shells we use the spin-unrestricted formulation of the HF and DFT theories.
In short, in our formulation, we perform  the spin-dependent average of the effective potentials within each sub-shell 
of the same $n$ and $l$, resulting in a single radial function. As we discuss in the next subsection, 
these radial functions are then directly imported into QMC methods.

In general, the Kohn--Sham and HF wave functions have a form of a single Slater determinant
and are explicit eigenstates of $M_S$, but they are eigenstates of $S^2$ only if $|M_S|$ has 
the maximum value. Therefore, the $L$ and $S$ eigenstates with maximum multiplicity ($2S+1$) 
and with the largest value of $L$ automatically have the correct spin symmetry. 

In order to simplify the DFT implementation, for open shell cases, we
have neglected the angular dependence of the spin-densities for
calculation of the exchange-correlation.  This treatment is commonly
known as the spherical approximation used in many DFT atomic solvers
for production of pseudopotentials. A very good numerical agreement (better than 1 mHa) was 
achieved between our and other atomic solvers (OPIUM~\cite{OPIUM}, FHI98PP~\cite{FHI98PP} and APE~\cite{APE}) for 
several open and closed shell atomic states and functionals. 
The four rungs of Perdew's ladder of approximate DFT functionals are represented by 
the Perdew--Wang (PW) LDA~\cite{PerdewWang92}, Perdew--Burke--Ernzerhof (PBE) GGA~\cite{PBE96}, 
Tao--Perdew--Staroverov--Scuseria (TPSS) meta-GGA~\cite{tpss} and PBE hybrid (PBE0)~\cite{PBE0}
exchange-correlation functionals as implemented in the LIBXC
library~\cite{libxc} and in the Quantum Espresso suite~\cite{qs}. 

\subsection{Quantum Monte Carlo Methods}
The trial many-body wave function serves as the most important input for the quantum Monte Carlo methods
within the fixed-node diffusion Monte Carlo approach. 
For this problem, following the approach used in previous
studies~\cite{Ballone92,Harris98,Sottile01}, we choose a
many-body wavefunction that is a product of spin-up and spin-down Slater determinants ($D$) 
and a Jastrow correlation factor ($J$), written as 
\begin{align}\label{eg:psi_qmc}
\Psi_T({\bf r}_1, {\bf r}_2, \ldots, {\bf r}_N)= &D[\varphi^{\uparrow}_1({\bf r}_1),\ldots,\varphi^{\uparrow}_M({\bf r}_M)]\nonumber \\
 &\times D[\varphi^{\downarrow}_{1}({\bf r}_{M+1}),\ldots,\varphi^{\downarrow}_{N-M}({\bf r}_N)] \nonumber\\
 &\times\exp[J({\bf r}_1, {\bf r}_2, \ldots, {\bf r}_N)],
\end{align}
assuming the first $M$ electrons to be spin-up and the remaining $N-M$ to be spin-down. The Slater determinants are constructed 
from orbitals generated in single-particle methods and have a form
\begin{align}
\varphi^{\uparrow(\downarrow)}_{k=n,l,m}({\bf r}_i)=R^{\uparrow(\downarrow)}_{n,l}(r_i)Y_{l,m}(\theta_i,\phi_i).
\end{align}
The symmetric Jastrow correlation factor includes well-known electron-electron cusp conditions as well as  
one and two-body correlation functions (for details see, e.g., Ref.~[\onlinecite{bajdichtheses}]). 
As a note, we did not find it necessary to include the multipolar terms into the Jastrow factor as in Ref.~\onlinecite{Sottile01}.

The Jastrow term is further variationally optimized~\cite{Umrigar05,Umrigar07}.
The optimal $\Psi_T$ is then used in diffusion Monte Carlo (DMC) method to obtain the ground state fixed-node (FN) energies and other 
expectation values of the system. As a additional step, for selected states, we also 
perform reptation Monte Carlo (RMC) calculations~\cite{rmc} to obtain pure expectation values of the fixed-node densities. 
All the QMC calculations were performed using QWalk code~\cite{Qwalk}.

As we have already mentioned, the $L$ and $S$ eigenfunctions considered in this paper with $M_L\neq 0$ are complex-valued
and require the use of the fixed-phase DMC algorithm\cite{fixedphase}. However, the associated fixed-phase errors are in general different 
(and possibly larger) than the fixed-node errors for real-valued wavefunctions\cite{FARcomplex}.
To avoid mixing results with two different approximations, we exclusively use the real-valued $M_L=0$ projections of 
the same $LS$ eigenfunction (as they are energetically degenerate). The $M_L=0$ projections can be readily obtained 
from $M_L=L$ states by the recursive application of the momentum lowering $L^-$ operator. As a consequence, 
the complex-valued single Slater determinant is replaced by the real-valued linear combination of Slater 
determinants. Please refer to the Appendix for the linear combinations
used for each open shell state.

\section{Results}~\label{results}
\subsection{Tests and validations}
To test the numerical implementation of our HF, DFT and fixed-node (FN) DMC methods
we have recalculated previous results for the closed-shell jellium spheres.
We achieve an excellent agreement with the PW LDA and PBE GGA energies of Ref.~\onlinecite{Almeida02}
and with TPSS meta-GGA energies of Ref.~\onlinecite{Almeida08}.
Obtained results for HF energies are identical to
Ref.~\onlinecite{HF200e} for model sodium clusters of $r_s=4.0$ with up to 196 electrons. 
However, we find lower HF energies (by $0.5$ mHa per electron) when compared to energies
of Ref.~\onlinecite{Sottile01}. This is most likely due to the
non-self consistent treatment of HF in Ref.~\onlinecite{Sottile01}. 
Finally, all our FN-DMC energies using LDA orbitals are within error-bars of FN-DMC energies of the Ref.~\onlinecite{Sottile01}. 
The sole exception is the 2 electron system at $r_s=1$, where the differences are larger.

The single limitation for the DMC calculations comes from the use of fixed-node approximation.  
Introduced fixed-node errors can be reduced by expanding the trial wave function in multi-determinants~\cite{Umrigar07, shdmcletter} or 
by using the backflow correlation corrections~\cite{Feynman56,Kwon98,Rios06,Bajdich08}, or both.
Previously, it was found that the QMC calculations for the closed shell jellium spheres\cite{Sottile01}
did not suffer from large fixed-node errors (by comparing to FN-DMC results with small multi-determinant expansions~\cite{Harris98}). 
In order to systematically check for these errors, we have extended the previous FN-DMC calculations to 
include backflow correlations for all the densities and particle
numbers. We find small and uniform gains to correlation energies on the order of 0.3 mHa per electron.
Therefore, it is very reasonable to assume that, for the energy differences considered here, these small corrections cancel out. 

\subsection{Results for the spherical jellium with impurity}
\subsubsection{$1d \to 2s$ transitions for N=18}
The excellent agreement for the closed-shell energies of jellium spheres, obtained with both with QMC and DFT methods, with earlier publications
 gives us confidence to study open shells.
As the first testing case, we have selected a jellium sphere with 18
electrons, which for the impurity free system is one of the 
fully-filled closed shells and corresponds to a cluster with a stable configuration.
As we increase the attractive Gaussian potential of the impurity, the nearest unoccupied $2s$ level is lowered in energy 
below the $1d$ level. Interestingly, exactly the same crossing of the
levels was previously observed in  the
context of hetero-atomic clusters~\cite{cohen87,Baladron88} and 3D quantum dots~\cite{vorrath2003}.
As a result, the ground state of the system will change from closed shell occupation $^1S(1s^21p^61d^{10})$ to 
either  $^3D(1s^21p^61d^{9}2s^1)$ or $^3F(1s^21p^61d^{8}2s^2)$ open shell occupations. As a note, each state differs in occupation 
in more than one shell ($1d$ and $2s$) while Hund's rules strictly apply only to occupations within a single shell.

\begin{figure}
\includegraphics[width=\columnwidth]{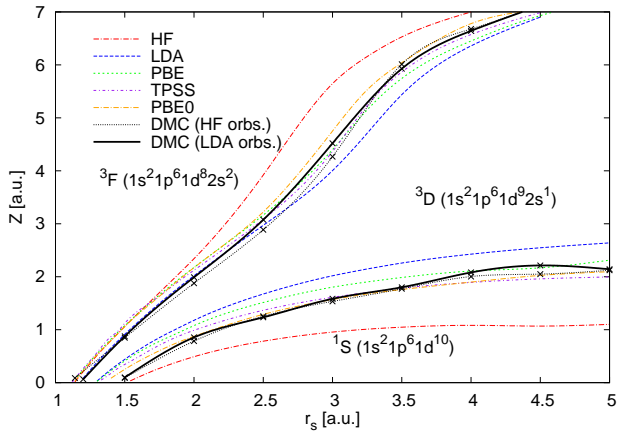}
\caption{Phase diagram of spherical jellium with 18 electrons and for fixed $\sigma=0.6$ as a function of
density $r_s$ and impurity potential strength $Z$. The transitions
between the three lowest lying states  
are shown for unrestricted Hartree--Fock (HF) (red dot-dashed), PW LDA (blue dashed), 
PBE GGA (green short-dashed), TPSS meta-GGA (purple dot-dashed), PBE0 hybrid (orange dot-dashed),
DMC with UHF orbitals (black dot-dashed) and DMC with LDA orbitals (black full) lines (see also text). 
The lines in the figure are extrapolations over the discrete points by cubic splines. }
\label{phase_space_d_shell}
\end{figure}

The phase diagram depends on the variables
 $r_s$, $Z$ and $\sigma$. 
We limit our 
discussion to realistic density range between $r_s=1$ and $r_s=5$ found in the majority of bulk systems.
Since the $\sigma$ and $Z$ parameters are effectively coupled together, we choose to fix the width of the Gaussian 
at $\sigma=0.6$ and vary only its depth $Z$. As a result, we find robust $1d \to 2s$ transitions within $0<Z<7$.
We summarize our calculations in the phase diagram,  Fig.~\ref{phase_space_d_shell}. 

There are several general observations which can be drawn from the phase diagram, Fig.~\ref{phase_space_d_shell}. 
As we increase the attractiveness of the impurity, we see changes on the occupation of a more delocalized $1d$ orbital
to more localized atomic-like $2s$ orbital in the following order: $1d^{10}\to 1d^{9}2s^1 \to 1d^{8}2s^2$.  
In addition, we notice the  existence of high density region ($r_s < 1.4 $) which has always 
partial $2s$ occupation (see also Table~\ref{correctiontable}). This region for jellium spheres at $r_s=1$ was not discussed 
 in previous studies. 
\begin{table}
\caption{Comparison between the total energies of the
closed-shell $^1S$ and open-shell $^3F$ states for $N=18$ and $r_s=1$.
The latter state is clearly lower in energy.}
\label{correctiontable}
\begin{tabular}{l c c c c c }
\hline
\hline
state &  E$^{DMC}$ &  E$^{LDA}$ &  E$^{PBE}$  &  E$^{HF}$ \\
\hline
$^1S$ &  0.39104(2) &  0.39317 &  0.38908 &  0.42786 \\
$^3F$ &  0.38817(2) &  0.39030 &  0.38628 &  0.42386 \\
\hline
\end{tabular}
\end{table}

From a methodological point of view, Fig.~\ref{phase_space_d_shell} provides 
detailed comparison between the single-particle methods and our DMC benchmarks. 
Our first observation is that HF greatly overestimates while LDA underestimates
the region of the stability of the  $^3D(1s^21p^61d^{9}2s^1)$ state. In fact LDA and HF bracket our best estimate, 
which is not surprising considering the success of hybrid functionals.  
Second, the GGA rung of the functional ladder represented by the PBE leads to clear improvement in accuracy over LDA. 
Next, the TPSS meta-GGA and PBE0 hybrid functionals agree even closer with DMC predictions than PBE
however without particular order in accuracy. 

Last, to give some measure of the fixed-node errors, 
we also compare the DMC results using LDA orbitals 
with DMC results using HF orbitals. Despite the different nodes
resulting from these choices the DMC energies are very similar
indicating that the nodal errors remain small.

\subsubsection{$1f \to 2p$ transitions for N=30}
In general, the behavior of the 30 electron system with a
partially-occupied $1f$ shell is similar to the case with 18
electrons.  The main difference is that, for the impurity free system, 
the ground state configuration is characterized by a 
high-spin quintet state at $1f$ shell and with $2p$ as the closest empty level.  
Therefore, the role of the extended orbital that transfers the electron is taken by $1f$ 
while $1p$ localizes and receives an electron as the impurity potential is applied.

Since $1p$ is more degenerate than $1s$, the 30 electron systems allows us to study
more complicated effects of correlations and exchange at the impurity as we change the impurity potential. 
The accessible states of the interest are  
$^5I(1s^21p^61d^{10}2s^21f^{10})$,  $^7I(1s^21p^61d^{10}2s^21f^92p^1)$, $^9G(1s^21p^61d^{10}2s^21f^82p^2)$ and
$^{11}S(1s^21p^61d^{10}2s^21f^72p^3)$. 

\begin{figure}\includegraphics[width=\columnwidth]{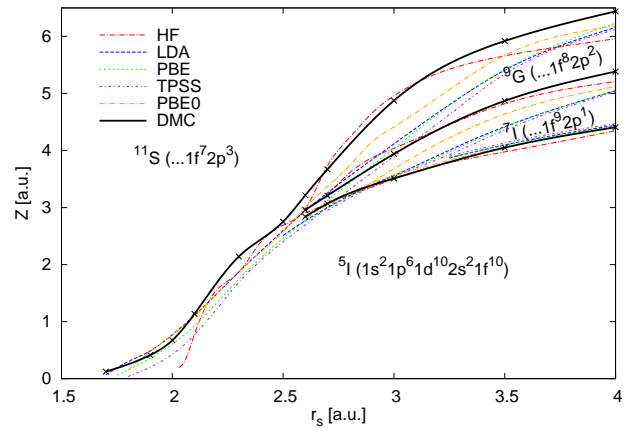}
\caption{Phase space diagram for jellium with 30 electrons and for fixed $\sigma=1.5$ as a function of
density $r_s$ and impurity strength $Z$. We have identified $^5I$, $^7I$, $^9G$ and $^{11}S$ states as 
occupying this section of the $r_s$-$Z$ space. The convention for lines is identical to Fig.~\ref{phase_space_d_shell}.}
\label{phase_space_f_shell}
\end{figure}

In Fig.~\ref{phase_space_f_shell} we directly compare the phase diagram obtained with DMC and mean field  methods
used in the 18 electron case. The detailed comparison shown in Fig.~\ref{phase_space_f_shell} reveals several important features. 
First,  LDA, PBE GGA and TPSS meta-GGA, all semilocal functionals, predict almost identical boundaries. 
When compared with our DMC results, we find that only the locations of $^5I \leftrightarrow ^{11}S$ and $^5I \leftrightarrow ^7I$  phase transitions agree well, 
while the  $^7I \leftrightarrow ^9G$ and $^9G \leftrightarrow ^{11}S$ transitions are shifted to lower $Z$ values. 
Second, PBE0 hybrid corrects slightly for the lower $Z$ shifts 
in the  $^7I \leftrightarrow ^9G$ and $^9G \leftrightarrow ^{11}S$ 
transitions.  In contrast, 
HF method produces much more satisfactory agreement with DMC for smaller densities ($r_s>2.5$) 
but greatly overestimates the stability of $^{11}S$ state at higher densities ($r_s<2.2$) due to the missing correlation.
The behavior described above suggest that a) the inhomogeneity effects in the density are relatively small 
and well captured by the semilocal functionals; b) the full non-local exchange absent at the semilocal level 
and only partially present (25\%) in PBE0 functional is needed to correct for lower $Z$ shifts. 


\subsection{Results for the spherical jellium shell with impurity}
\subsubsection{$1d \to 2s$ transitions for N=18}
As in the first case of the spherical jellium with impurity, we choose to study the 18 electron system in the spherical 
shell potential [Eq.~(\ref{eq:vextshell})]. 
The occupation of the single-particle states for hollow cluster is assumed to be the same as for jellium spheres 
(as confirmed by Ref.~\onlinecite{hollowclusters09}). Rather than finding the extensive phase space diagram 
of the system we limit our study to a very small subset of the space. 
We select $M=9$, $r_s=3.0$ and $Z\sim 8.5$ to illustrate a size of localization errors from HF and DFT based theories. 

\begin{figure}
\includegraphics[width=\columnwidth]{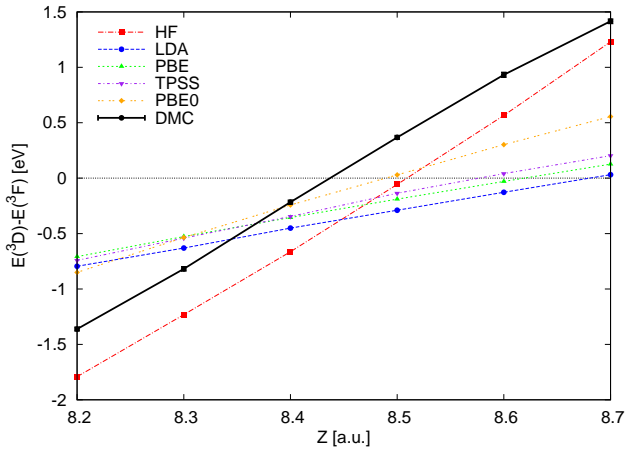}
\caption{Total energy difference between $^3D$ and $^3F$ states of the spherical jellium shell 
of 18 electrons with $M=9$, $r_s=3.0$ and $\sigma=0.6$ in a small window of $Z$. 
All the methods are described in the text. }
\label{fig:de_shell}
\end{figure}

Figure~\ref{fig:de_shell} compares the energy difference between $^3D$ and $^3F$ states as a function of $Z$ 
calculated with single-particle methods and DMC. 
Arguably, the main point of interest in Fig.~\ref{fig:de_shell} is the slope around $^3D \leftrightarrow ^3F$ transition
and secondly the size of the relative errors.  As in previous subsection, all semilocal functionals (i.e., LDA, PBE, TPSS) 
results have a very similar slope (with half the steepness of the DMC curve). 
On the other hand, the slope of the HF curve  is almost identical to DMC curve. Not surprisingly, 
PBE0 partially recovers the correct slope as it contains 25\%  of exact exchange. 

\begin{figure*}
\begin{tabular}{c c}
\includegraphics[width=0.5\textwidth]{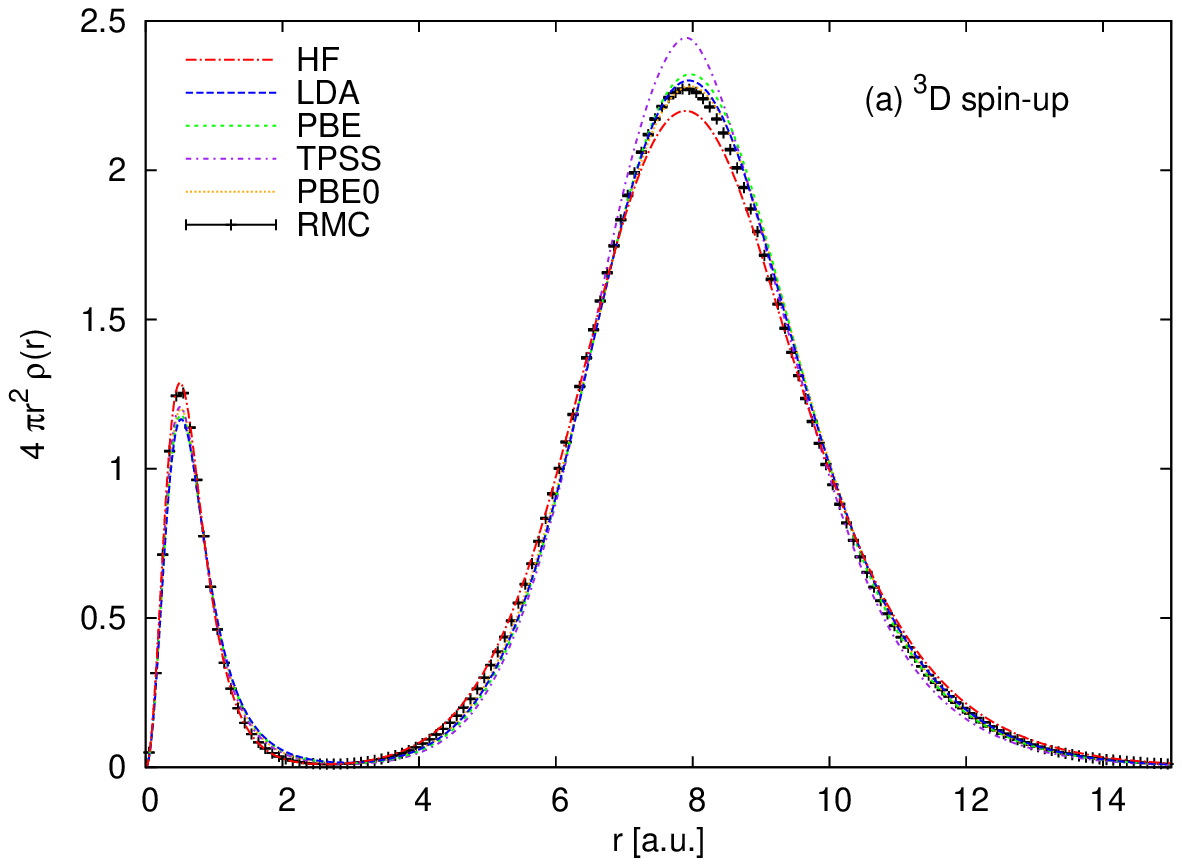} & 
\includegraphics[width=0.5\textwidth]{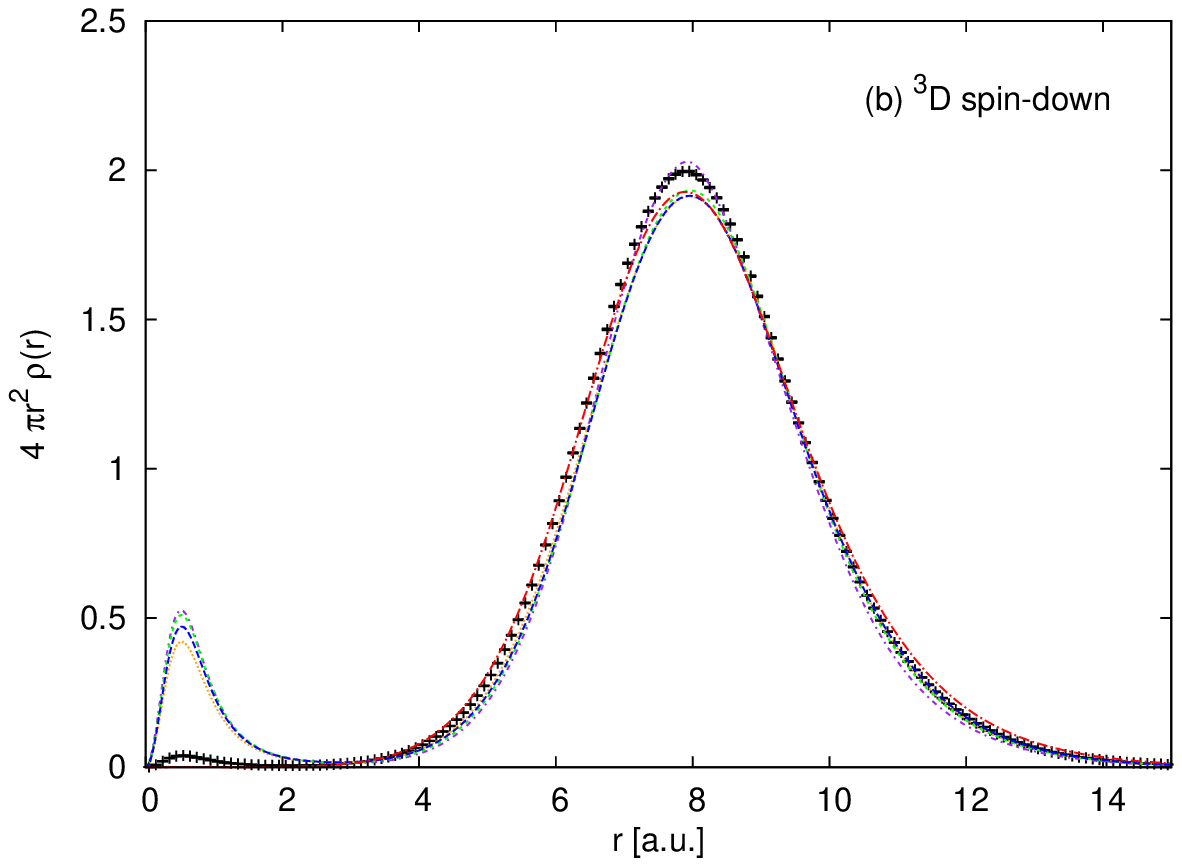} \\
\includegraphics[width=0.5\textwidth]{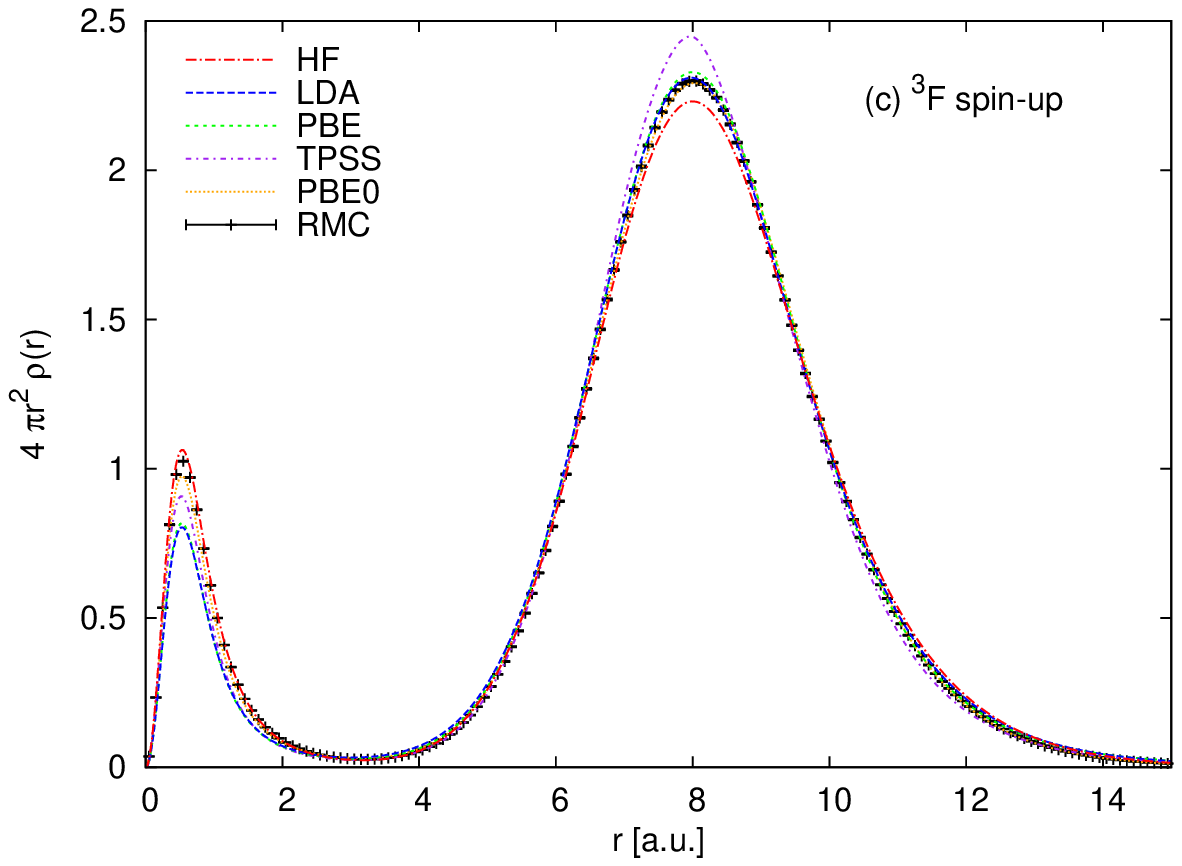} & 
\includegraphics[width=0.5\textwidth]{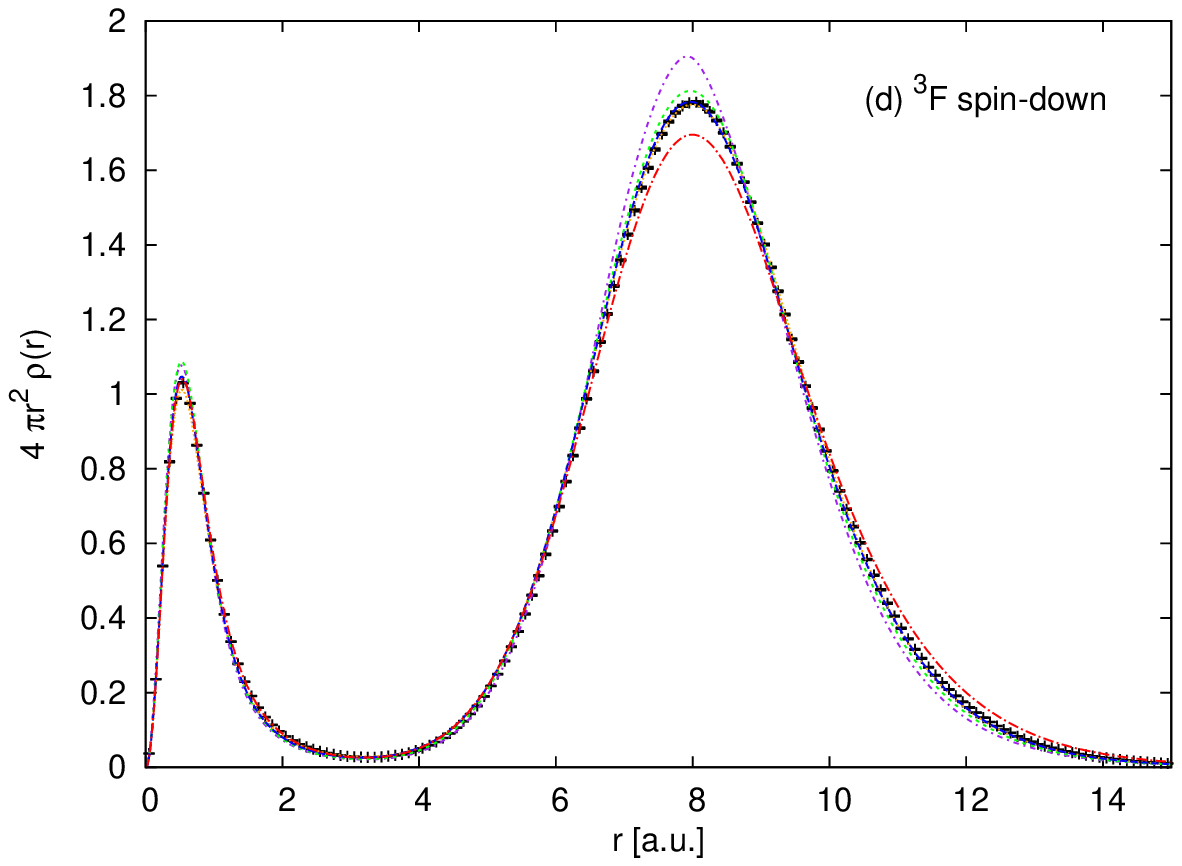} \\
\end{tabular}
\caption{Radial densities for the spherical jellium shell of 18 electrons for $^3D$ (upper) $^3F$ (lower) states 
with $M=9$, $r_s=3.0$, $Z=8.5$ and $\sigma=0.6$. Spin-up channel (left figure) and 
spin-down channel (right figure). All the methods are mentioned in the text and the 
convention for lines is identical to Fig.~\ref{phase_space_d_shell}.
The pure expectation of the density operator from RMC method employed LDA orbitals.}
\label{fig:shell_density}
\end{figure*}

Finally, it is also instructive to analyze the radial densities for the $^3D$ and $^3F$ states (see Fig.~\ref{fig:shell_density}).
As a benchmark we use the pure expectation of the density operator from reptation Monte
Carlo~\cite{rmc}(RMC) method with LDA orbitals.  
The densities for each state and spin channel in
Fig.~\ref{fig:shell_density} have a distinct double peak structure --
the smallest
is due to the presence of the Gaussian impurity and the largest due to the spherical shell itself. 
Also visible is the relative reduction of the smaller peak for the $^3D$  spin-down channel due 
to the absence of the more localized $2s$ state. 

From the direct comparison with RMC results we deduce that the density at inner shell regions (i.e., smaller peak) 
is better described within HF while LDA and PBE GGA provide better densities for the outer shell regions (i.e., larger peak).
The PBE0 hybrid smoothly interpolates between PBE and HF limits with
the best overall description. 
Lastly, the TPSS meta-GGA results in an substantial increase of density at both peaks and decrease at the tails
when compared to  LDA and PBE GGA. We find this overcompensation of
the density, presumably due to the 
kinetic energy density contributions specific to the TPSS functional, to be surprisingly high.


\section{Summary}
In conclusion, we have studied the performance of 
a variety of DFT functionals with increasing complexity  
against quantum Monte Carlo benchmark in a spherical jellium model and an spherical shell model with 
an attractive Gaussian impurity at the center.
The tunable strength of the impurity allowed us to find a number of interesting transitions 
between closed and open shell states. Our results shows that the development of better density functional approximation is
increasingly required as the system departs from the perfect spherical jellium case. 
We report several regions where the employed approximations of DFT fail to find 
the same ground state as identified with QMC methods. 

The $1d \to 2s$ transitions in 18 electron system are well 
described on the highest semilocal level (TPSS meta-GGA) as well in global hybrid (PBE0) DFT.
On the other hand, not all $1f \to 2p$ transitions in the 30 electron system 
were accurately captured. We argue that even as the inhomogeneity effects in the density are relatively small 
and well captured by the semilocal functionals, full non-local exchange is needed to accurately describe the system.
Our work therefore further supports the need for the hyper-GGA functionals~\cite{hyperGGA} with 
fully non-local exchange and accompanying balanced correlation.

In the spherical jellium-shell model with an impurity at the center, where the inhomogeneity in the electronic density is increased,  
the DFT methods with exact-exchange give better agreement  for the studied transitions. 
The radial electron densities in the inner region closest to the
impurity are correctly described at the HF level, while LDA and PBE GGA are more accurate in the outer region.
The PBE0 global-hybrid results smoothly interpolates between HF and GGA. We find surprisingly high deviations in density 
for the TPSS.  Finally, we also publish our results in the EPAPS Document No. [] with the purpose of allowing a detailed comparison 
with newly developed DFT functionals.

\section*{Acknowledgments}
The authors thank Markus D\" ane, Markus Eisenbach, Don M. Nicholson  
and G. Malcom Stocks for their contributions at the early stages of
this project and acknowledge Valentino R. Cooper's careful reading of the manuscript.       
M.B. would also like to thank to Xiong Zhuang for access to his $LS$ eigenfunction program.
This research used computer resources supported by the U.S. DOE Office of Science under
contract DE-AC02-05CH11231 (NERSC) and DE-AC05-00OR22725
(NCCS). Research sponsored by U.S. DOE BES Divisions of Materials
Sciences \& Engineering (FAR) and Scientific User Facilities
(PRCK), and the ORNL LDRD program (MB).

\section*{Appendix: Real-valued $LS$ eigenfunctions}\label{appendix}
Application of $(L^-)^L$ operator on the $LS$ eigenfunction with 
$M_L=L$ leads to real-valued $LS$ eigenfunction with $M_L=0$.
The linear combinations of Slater determinants for the open-shell states in 18 electron system are then
\begin{align}
\Psi[^3D(1d^{4\downarrow})]=& D_1\\
\Psi[^3F(1d^{3\downarrow})]=& \frac{1}{\sqrt{5}}(D_2+2D_3),
\end{align}
where $D_1=(2^-,1^-,-1^-,-2^-)$, $D_2=(1^-,0^-,-1^-)$ and $D_3=(2^-,0^-,-2^-)$ are determinants 
with indicated relevant occupied orbitals (numbers stand for the orbital quantum numbers and superscripts indicate the spins).

The $M_L=0$ linear combinations for the 30 electron system are 
\begin{align}
\Psi[^9G(1f^{\downarrow}2p^{2\uparrow})]=&\frac{1}{\sqrt{7}}\left[\sqrt{2}(D_1 +D_2) +\sqrt{3} D_3\right],
\end{align}
where $D_1=(-1^-,1^+,0^+)$, $D_2=(1^-,0^+,-1^+)$, $D_3=(0^-,1^+,-1^+)$ and
\begin{align}
\Psi[^7I(1f^{2\downarrow}2p^{\uparrow})]=&\frac{1}{6\sqrt{7}\sqrt{11}} [ 30D_4+24D_5+6D_6 \nonumber \\
          &+5\sqrt{6}D_7+9\sqrt{5}D_8+5\sqrt{3}D_9 \nonumber \\
          &+5\sqrt{6}D_{10}+9\sqrt{5}D_{11}+5\sqrt{3}D_{12}],
\end{align}
where $D_4=(1^-,-1^-,0^+)$, $D_5=(2^-,-2^-,0^+)$, $D_6=(3^-,-3^-,0^+)$, 
$D_7=(0^-,-1^-,1^+)$, $D_8=(1^-,-2^-,1^+)$, $D_{9}=(2^-,-3^-,1^+)$, 
$D_{10}=(1^-,0^-,-1^+)$, $D_{11}=(2^-,-1^-,-1^+)$, $D_{12}=(3^-,-2^-,-1^+)$ and
\begin{align}
\Psi[^5I(1f^{3\downarrow})]=&\frac{1}{\sqrt{2}\sqrt{3}\sqrt{7}\sqrt{11}} [ 5D_{13}+16D_{14}+9D_{15} \nonumber \\
          &+5\sqrt{2}(D_{16}+D_{17})],
\end{align}
where  $D_{13}=(1^-,0^-,-1^-)$, $D_{14}=(2^-,0^-,-2^-)$, $D_{15}=(3^-,0^-,-3^-)$, $D_{16}=(2^-,1^-,-3^-)$,$D_{17}=(3^-,-1^-,-2^-)$.
Above results have been also verified numerically using code from Ref.~\onlinecite{Zhuang}.

\end{document}